\documentstyle[11pt,epsf]{article}
\begin{document}
\title{\bf Testing causality violation on spacetimes with closed timelike 
curves}
\author{Seth Rosenberg\\ Department of Physics\\ University of California at
Santa Barbara, Santa Barbara, CA 93106\footnote{e-mail address: 
seth@physics.ucsb.edu}\\  }
\date{UCSBTH-97-17  \\ \today}
\maketitle
\def\< {\langle}
\def\> {\rangle}
\def\l {\lambda}
\def\G {\gamma}
\def\e {\epsilon}
\def\s {\sigma}
\def\o {\Omega}
\def\be {\begin{equation}}
\def\ee {\end{equation}}
\def\ba {\begin{eqnarray}}
\def\ea {\end{eqnarray}}
\def\dcs {\left(\frac{d\sigma}{d\Omega}\right)}
\def\g() {g(\psi_i,\phi)}
\def\ng() {g(\psi,\phi)}
\def\amp {{\left|\< \psi_i|\phi\> \right|}^2 }
\def\namp {{\left|\< \psi|\phi\> \right|}^2 }
\def\yint {\int_{-\frac{L}{2}}^\frac{L}{2}}
\def\half {\frac{1}{2}}
\begin{abstract}
	Generalized quantum mechanics is used to examine a simple
two-particle scattering experiment in which there is a bounded region
of closed timelike curves (CTCs) in the experiment's future. The
transitional probability is shown to depend on the existence and
distribution of the CTCs.  The effect is therefore acausal, since the
CTCs are in the experiment's causal future.  The effect is due to the
non-unitary evolution of the pre- and post-scattering particles as
they pass through the region of CTCs.  We use the time-machine
spacetime developed by Politzer~\cite{politzer}, in which CTCs are
formed due to the identification of a single spatial region at one
time with the same region at another time. For certain initial data,
the total cross-section of a scattering experiment is shown to deviate
from the standard value (the value predicted if no CTCs existed).  It
is shown that if the time machines are small, sparsely distributed, or
far away, then the deviation in the total cross-section may be
negligible as compared to the experimental error of even the most
accurate measurements of cross-sections.  For a spacetime with CTCs at
all points, or one where microscopic time machines pervade the
spacetime in the final moments before the big crunch, the total
cross-section is shown to agree with the standard result (no CTCs) due
to a cancellation effect.
\end{abstract}

\vfill

\pagebreak
\section{Introduction}
	Standard quantum theories consist of a Hilbert space of states
defined on spacelike surfaces and a unitary operator which evolves
these states through time.  The state may be the wave-function of a
particle, as in non-relativistic quantum mechanics, or the quantum
state of a matter field, as for quantum field theory.  When we
quantize gravity, we expect the spacetime metric, and therefore the
causal structure of spacetime, to vary quantum mechanically.  The
metric may be in a mixed quantum state, not corresponding to any
classical spacetime.  It will therefore be impossible to define
whether two points in the spacetime are spacelike, null, or timelike
separated, or to define a spacelike surface.  If we cannot define
states on spacelike surfaces, then we also lose any notion of the
unitary evolution of these states.  Even if there exist regions of
spacetime which are foliable by spacelike surfaces, we cannot say that
the operator which evolves states between two spacelike surfaces will
be unitary.

	Given this, the current framework for quantum theory will not
be sufficient to express a quantum theory of gravity.  What is needed
is a theory which can be expressed covariantly, without dependence on
spacelike surfaces.  One possibility is to follow Feynman and use a
sum-over-histories approach.  The fundamental constituents of
the theory are histories of alternatives, which are parameterized
paths of the given system through its phase space.  There is no
concept of evolution, since there are no states to evolve.  One may,
however, be able to assign probabilities to the histories if the
quantum interference between histories is negligible.  If so, then we
can calculate transition amplitudes between two different
configurations of the system by computing a path integral over all
possible histories between the two configurations~\cite{gqm}.

	Quantum mechanics also fails when applied to spacetime with
closed timelike curves (CTCs). These spacetimes are generally not
foliable by spacelike surfaces. Clearly the global causal structure
breaks down due to the CTCs, and there may be both spacelike and
timelike geodesics between two points in the spacetime.  Also, many
authors~\cite{politzer,friedman,boulware} have shown in perturbation
analyses that interacting theories exhibit non-unitary evolution
through a region of CTCs.  Using quantum theory in a spacetime with
CTCs one loses both the notion of a state on a spacelike surface and
the unitary evolution of those states.  These problems are similar to
those discussed above which occur when trying to quantize
gravity. Thus, spacetimes with CTCs may be useful as models in the
search for a quantum theory of gravity, since they provide simple
examples of spacetimes where quantum mechanics breaks down due to the
loss of causal structure. We can apply generalized quantum mechanics
to these spacetimes to see how this expanded framework handles the
breakdown of causal structure and the loss of unitarity.

	  One might expect that if the region of CTCs was bounded in
time, then the covariant theory of generalized quantum mechanics might
reduce to the standard causal theory before the region of CTCs.
However, in Hartle's~\cite{JimCTC} formulation of a prescription to
apply generalized quantum mechanics to a spacetime of this type, he
finds the theory to be acausal \footnote{Anderson applies a
less-straightforward prescription and finds no
acausality~\cite{anderson}}: in addition to the expected acausality
due to the presence of the CTCs, he also finds that if the region of
CTCs is bounded in time, alternatives occurring before the region of
CTCs will be affected by the existence and distribution of the later
CTCs.

	The acausal effect is evident when the total cross section of
a simple two-particle scattering experiment is calculated using
generalized quantum mechanics.  The cross-section has different values
depending on the existence and configuration of CTCs in the
experiment's future.  We will calculate this effect for a simple case
and discuss the size of the effect with different CTC configurations.

	In section 2 we find an expression for the probability for a
transition from one state to another given a bounded region of CTCs to
the future of the experiment.  The region of CTCs is bounded in time,
and so the regions of spacetime before and after the CTCs will be
foliable by spacelike surfaces.  Therefore, we will be able to define
states on spacelike surfaces and unitary evolution outside the CTC
region.  The transition probability is shown to be explicitly acausal
due to its dependence on the non-unitary evolution caused by the
future CTCs.  In section 3 we first specify the background spacetime
for our calculation: a flat, non-relativistic spacetime, with an
identification made between one spatial region at one time and the
same spatial region at another time. This spacetime was introduced by
Politzer~\cite{politzer}, and has been studied by many
authors~\cite{cornhole,hawking,mike}.  This identification creates a
region of CTCs between the two identified times.  In order to
calculate the probability derived in section 2, we first examine the
non-unitary evolution of the pre- and post-transition states through
the region of CTCs.  We arrive at a rough estimate for this effect by
calculating the non-unitary evolution as a function of the spatial
distance between a Gaussian wave-function and the center of the
identified region.  This is done using the Born approximation in
non-relativistic quantum mechanics with a contact potential.  The
non-unitary evolution occurs as a single particle passes through the
region of CTCs and scatters off of a future version of itself or off
of a particle trapped inside the time machine.

	In section 4 we calculate the deviation this effect causes in
the total cross-section of a two particle scattering experiment if
CTCs exist in the future of the experiment.  If neither pre-scattering
particle is aimed at the identified region, we find that for a specific
range of scattering energies, the total cross-section will deviate
from the value predicted if no CTCs existed.  This deviation is shown
to depend on a number of factors, including the total solid angle
subtended by the time machine as viewed from the scattering center.
It is shown that if the time machines are small, sparsely distributed,
or far away, then the deviation in the total cross-section may be
negligible as compared to the experimental error of known cross-
sections. Limits are placed on the possible distributions of identified
regions given known errors of scattering experiments.  For a
spacetime filled with microscopic Politzer identifications, or with an
isotropic distribution of these identifications, the cross-section is
shown to agree with the standard result (with no CTCs) due to a
cancellation effect.
 
\section{The effect of future CTCs on alternatives in generalized
	     quantum mechanics}

	We wish to explore the acausality discussed in the
introduction.  Hartle's prescription for applying generalized quantum
mechanics to spacetimes with CTCs produces an acausal
theory~\cite{JimCTC}.  This acausality can be seen by examining a 
two-particle scattering experiment with a region of CTCs after the
experiment.  The total cross-section of the scattering experiment will
be different from the value predicted if CTCs were not present.  To
describe this effect, we begin by applying generalized quantum
mechanics to a quantum transition when a bounded region of CTCs exists
in the future of the transition.

	We will use a fixed background spacetime which is flat and
non-relativistic everywhere with a region of CTCs which begins at time
$t_-$ and ends at $t_+$.  The transition of interest will be a two
particle scattering event which both begins and ends before $t_-$.
The time evolution operator outside the time interval $[t_-,t_+]$ will
be the standard time evolution operator $U(t',t)$ between two times
$t$ and $t'$.  We cannot express the evolution of states inside
$[t_-,t_+]$ as an operator, since states are not defined in this
region of spacetime.  However, we will express the evolution from
$t_-$ to $t_+$, completely through the region of CTCs, as an operator
$X_s$, which may not be unitary.

	We split our Hilbert space into a subspace which covers only
the particles involved in the quantum transition, and another subspace
which covers the remaining variables.  We will call these two
subspaces ``the experiment'' and ``the environment'' respectively. We
assume that these subspaces are only weakly coupled.

	In the language of generalized quantum mechanics, we begin
with an initial density matrix $\rho$ at time $t_1$, and define a set
of alternatives for the post-scattering state at time $t_2$.  Since
$t_2$ is before the region of CTCs, we may express these alternatives
as a set of projection operators ${{P_\alpha}}$ at time $t_2$.  Each
coarse grained history will be identified with a given $P_\alpha$.
Using the Heisenberg picture, the projection operators take the form
\be \label{Hproj}
P_\alpha(t_2) =  U^{-1}(t_2,t_1) P_\alpha U(t_2,t_1) \; . 
\ee
The transition of interest is a two-particle scattering experiment,
and so our alternatives will be ranges of position and momentum for
both particles in the post-scattering state.  This scattering
experiment and its background spacetime are illustrated in figure 1.

\begin{figure}
\leavevmode
\centering
\epsfysize=15cm 
\epsfbox{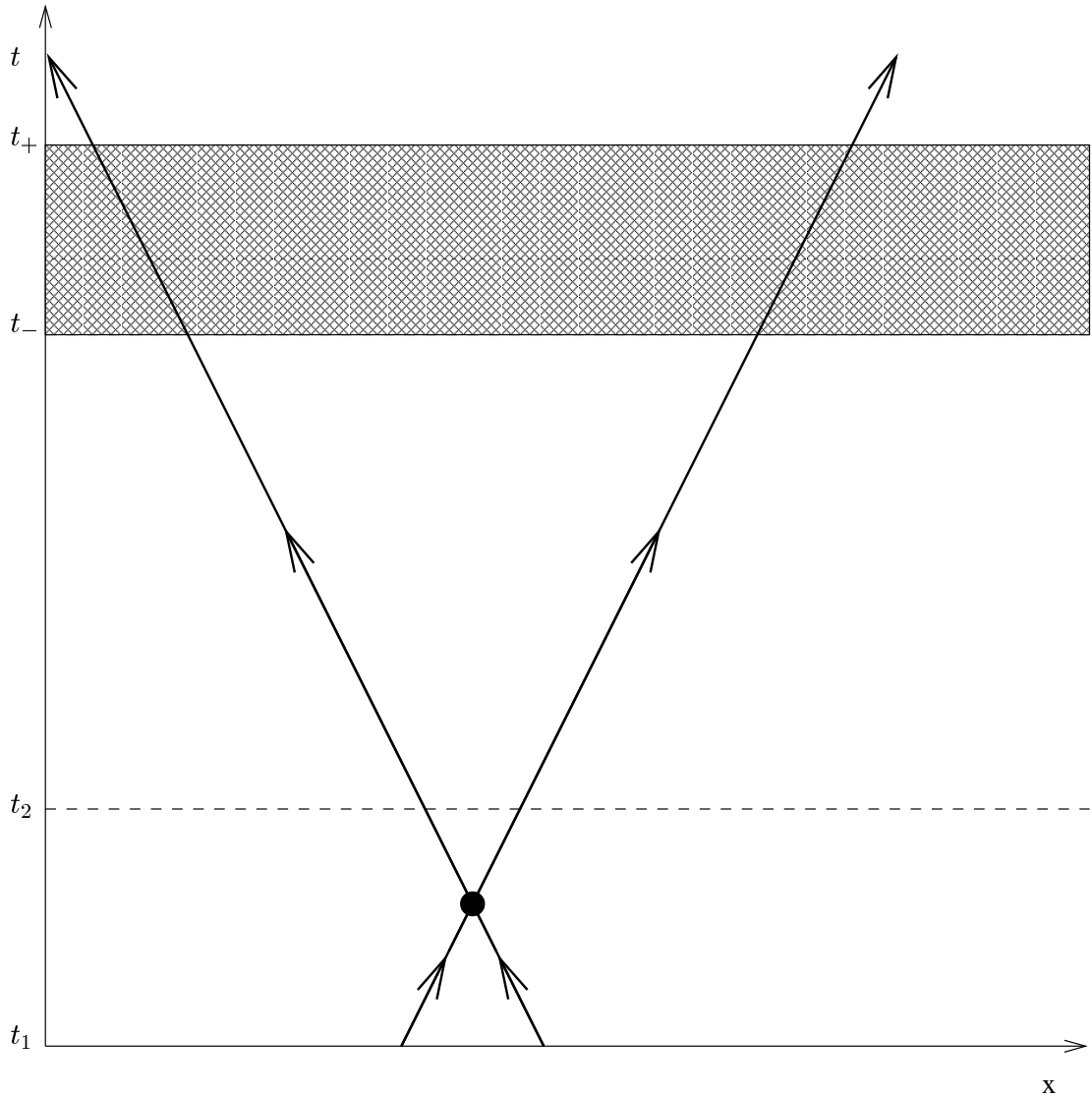}
\caption{The scattering event and its background spacetime.  The event
takes place between $t_1$ and $t_2$, and the region of CTCs is between
$t_-$ and $t_+$.
\label{fig1}}
\end{figure}

In generalized quantum mechanics it is not always possible to assign a
probability to a given history. Two coarse-grained histories are said
to decohere if the quantum interference between the two histories is
zero.  This quantum interference is measured by the decoherence
functional, which is a complex functional of two coarse-grained
histories. Probabilities can be defined for a given exhaustive
coarse-graining only if every pair of histories in the coarse-graining
decohere~\cite{gqm}.  For this paper, we will assume that the weak
coupling of ``the experiment'' to ``the environment'' will cause
decoherence, and therefore all probabilities are well
defined~\cite{JimMurray}.

Since our probabilities are well defined, we may use generalized
quantum mechanics to calculate the probability of a given
coarse-grained history ~\cite{JimCTC}.  We do so by stringing together
a series of Heisenberg operators, as defined in (\ref{Hproj}), which
act on our initial density matrix $\rho $.  We simply insert the
non-unitary operator, $X$, into the string according to when the
region of CTCs occurs with respect to the various alternatives.  Thus,
the probability of an alternative $\alpha_i $ occurring when our
initial density matrix is $\rho $ is given by
\begin{equation} \label{genprob}
p(\alpha_i)= \frac{Tr[XP_{\alpha_i}\rho P_{\alpha_i}X^+]}{Tr[X\rho X^\dagger]}
\; ,
\end{equation}
where
\[ X = U^{-1}(t_+,t_f)X_s U(t_-,t_1) \]
and $t_f$ is the final time, to the far future, where we trace over
all final states.  Equation (\ref{genprob}) expresses the probability
of a system beginning in a initial configuration represented by
$\rho$, and at some point in the history, one or more variables
which represent the system falling into a range corresponding to an
alternative $\alpha_i$. The latter part is represented by the
projection operator $P_{\alpha _i}$.  The system then experiences
non-unitary evolution due to the CTCs, and finally we sum over all
possible final configurations.

The probability for a given history depends on the non-unitary
evolution of the particles after the conclusion of the experiment.
Thus, the theory is manifestly acausal.  If there is no region of CTCs
to the future of the experiment, the probability reduces to
\begin{equation}
p(\alpha_i) = Tr[P_{\alpha_i} \rho P_{\alpha_i}] \; ,
\end{equation}
which is causal.  We will now find the deviation in $p(\alpha_i)$ if
CTCs exist in the future of the experiment.

Since both $t_1$ and $t_2$ are before $t_-$, we can express $\rho$ and
$P_{\alpha_i}$ in terms of quantum states at these times.  For
simplicity, we assume pure states:
\begin{eqnarray}
\rho \! \! \! & = & \! \! \!   |\phi\> \< \phi| \; ; \\
P_{\alpha_i} \! \! \!  & = & \! \! \! |\psi_i\> \< \psi_i| \; ; \nonumber
\end{eqnarray}
where $|\phi\> $,$|\psi_i\> $ are two-particle quantum states defined
at $t_1,t_2$ respectively.  Substituting these into (\ref{genprob}) we
have
\begin{equation} \label{newprob}
p(\psi_i) =  N \, Tr[X{|\psi_i\> }{\< \psi_i|\phi\> }{\< \phi|\psi_i\> }{\< 
\psi_i|}X^\dagger] \; ,
\end{equation}
where
\[ \frac{1}{N} = Tr[X|\phi\> \< \phi|X^\dagger] \; . \]
We now explicitly write the probability in terms of single particle
states.
\begin{eqnarray}
|\psi_i\> \! \! \! &=& \! \! \! |\psi_{i1}\> \otimes |\psi_{i2}\> 
\label{gfactor} \; ,\\
|\phi\> \! \! \! &=& \! \! \! |\phi_{1}\> \otimes |\phi_{2}\> \; . \nonumber
\end{eqnarray}
We will assume that $X$ can be factored into two non-unitary operators
$X_1$ and $X_2$, which evolve particles 1 and 2 respectively from
$t_-$ to $t_+$:
\begin{equation} \label{Xfactor}
X = X_1 \otimes X_2 \; .
\end{equation}
This assumption is sensible as we expect the distance between
particles after the scattering event to be large compared to the
interaction range.  Substituting (\ref{gfactor}) and (\ref{Xfactor})
into (\ref{newprob}), we have
%
%
%
%
\begin{equation} \label{longprob}
P(\psi_i) = N {|\< \psi_i|\phi\> |}^2 \< \psi_{i1} 
|{X_1}^\dagger X_1|\psi_{i1} \>  \< \psi_{i2}| 
{X_2}^\dagger X_2|\psi_{i2} \>  
\end{equation}
Many authors~\cite{politzer,friedman,boulware} have shown that the
operator which evolves a state from before a region of CTCs to after
that region is non-unitary for interacting theories in perturbation
theory. These authors showed that the non-unitary component of the
evolution is due to particle interactions in the appropriate theory.
When the evolution is expanded using perturbation theory, with a small
parameter $\l $ associated with the interaction, the non-unitary
component is of order $\l $ or higher.  Thus, we write
\begin{equation} \label{Xexpansion}
X_j^\dagger X_j = 1 + \lambda \hat{\epsilon}_{j_1} + \lambda^2  
\hat{\epsilon}_{j_2} + ... 
\end{equation}
where $j=(1,2)$. The $\epsilon_{j_n}$ are operators which express the
non-unitary component of the evolution, and $\lambda$ is some small
parameter.  We will only examine the evolution to first order, so
\begin{equation} \label{Xexpansion2}
X_i^\dagger X_i \approx 1 + \lambda \hat{\epsilon}_{i_1} \equiv 1 + \lambda
\hat{\epsilon}_i \; ,
\end{equation}
To first order in $\l $,
\begin{equation}
P(\psi_i) = N {|\< \psi_i|\phi\> |}^2 (1+\l \< \psi_{i1}|{\e}_1 |
\psi_{i1}\> )(1+\l \< \psi_{i2}|{\e}_2 |\psi_{i2}\> ) \; ,
\end{equation}
where
\begin{equation}
N^{-1} =  (1+\l \< \phi_1|{\e}_1 |\phi_1\> )
(1+\l \< \phi_2|{\e}_2 |\phi_2\> ) \; .
\end{equation}
Ignoring all terms of $o({\l}^2)$ or higher, we get
\begin{equation} \label{prob}
P(\psi_i) = {|\< \psi_i|\phi\> |}^2 \left[ \frac{1+\l ( \< \psi_{i1}|
{\e}_1|\psi_{i1}\> +  \< \psi_{i2}|{\e}_2 |\psi_{i2}\> ) } {1+\l ( \< \phi_1|
{\e}_1 |\phi_1\> +  \< \phi_2|{\e}_2 |\phi_2\> ) } \right] \; .
\end{equation} 
We see that the standard transition amplitude is modified by a
multiplicative factor.  This factor depends on the non-unitary
evolution that both the final and initial state experience as they
pass between $t_-$ and $t_+$.\footnote{For the initial state this
means the non-unitary evolution that the state would experience if it
evolved to $t_-$ without scattering.}

As noted before, if no CTCs exist in the experiment's future,
(\ref{prob}) reduces to
\be
p(\psi_i) =  {|\< \psi_i|\phi\> |}^2 \; .
\ee
This is the result predicted by standard quantum mechanics, which is causal.

\section{The position dependence of $\< \xi|\e_i| \xi\> $}

Equation (\ref{prob}) expresses the probability for a certain
alternative to occur after a scattering event.  Since this result
depends on the non-unitary evolution that both the pre- and
post-scattering states experience between $t_-$ and $t_+$ we must
investigate this effect in detail in order to calculate the total
cross section for a scattering experiment.  The post-scattering
trajectory of the particles, and their location at $t=t_-$ will depend
on the scattering angle. Since the identified regions are spatially
bounded, the non-unitary effect on a given particle may depend on the
spatial distance between the particle and the identified region as the
particle passes from $t=t_-$ to $t=t_+$. We let $|\xi \> $ represent a
generic particle which, in our case, will be one of the particles from
the scattering experiment.  We will find the behavior of the
non-unitary evolution which $|\xi \> $ experiences as a function of
the particle's distance from the center of the identified region.

There are many different spacetimes which contain CTCs.  For our
calculation we will use a 1+1 dimensional spacetime introduced by
Politzer~\cite{politzer}.  This spacetime is simple, yet has all the
essential features of a spacetime with CTCs.  We review the Politzer
spacetime, which is shown in figure 2.  The horizontal lines of length
$L$, centered at $z=0$, at $t=t_-$ and $t=t_+$ are identified so that
along them the region immediately before $t=t_-$ connects smoothly to
that after $t=t_+$, while the region immediately before  $t=t_+$
connects smoothly to that after $t=t_-$. We define $T \equiv t_+
-t_-$, $Y \equiv (-\frac{L}{2}, \frac{L}{2})$, and $R \equiv Y \otimes
(t_-,t_+)$.  We will refer to the identified regions as the time
machine, and $R$ as the interior of the time machine.  We will choose
the spacetime to be non-relativistic, and flat everywhere, with the
identification mentioned above.  This spacetime contains a region of
CTCs infinite in spatial extent, and bounded in time by $t=t_-$ and
$t=t_+$.

\begin{figure}
\leavevmode
\centering
\epsfysize=15cm 
\epsfbox{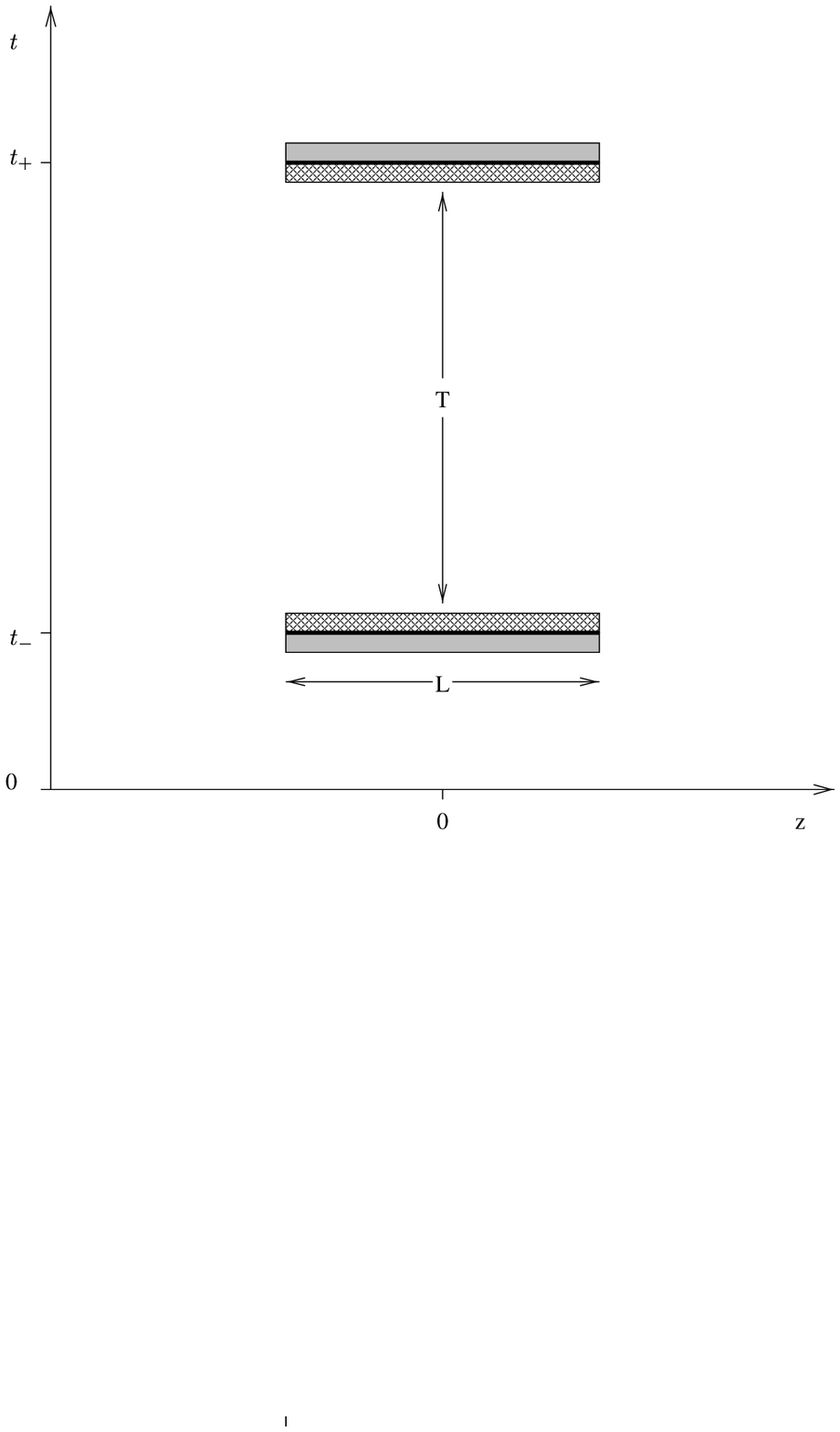}
\caption{The Politzer time-machine spacetime in 1+1 dimensions.  The
two shadings represent the identification between the two horizontal
lines at $t=t_-$ and $t=t_+$, so that $t<t_-$ connects smoothly with
$t>t_+$ and $t<t_+$ connects smoothly with $t>t_-$.
\label{fig2}}
\end{figure}

{\it Important notational convention}: Throughout this section we will
use the same position notation as Politzer~\cite{politzer}: $y$
variables refer to positions between $-{\frac{L}{2}}$ to
$\frac{L}{2}$, while $x$ variables refer to positions outside that
region and $z$ variables are unlimited.

After the scattering experiment the particles will evolve through
time until $t=t_-$ at which point the particles will pass through the
region of CTCs.  In this region each particle may interact with a
future version of itself or with a particle trapped inside the time
machine.  We expect that the inter-particle interaction
will be short ranged such as a screened Coulomb force or a nuclear
force.  Thus we model the interaction with a contact potential
$V(z,z') =\l V_0 \delta(z-z')$. Although the scattering experiment
itself will be considered in 3+1 dimensions in section 4, here we will
look at the interaction which occurs as a single particle passes by
the time machine in 1+1 dimensions for simplicity, and then generalize
our result to 3+1 dimensions.  This particle will be one of the pre-
or post-scattering particles, which has evolved through time up to
$t=t_-$. We will use bosons in non-relativistic quantum mechanics,
with the standard, non-interacting, flat space propagator:
\begin{equation}
K_f(z_2,t_2;z_1,t_1) = \left\{\begin{array}{ll}
{\left(\frac{m}{2\pi ih(t_2-t_1)}\right)}^{\frac{1}{2}} \exp
\left[\frac{im{(z_2-z_1)}^2}{2\hbar (t_2-t_1)}\right] & t_2 >t_1  \\
\delta (z_2-z_1) &  t_2=t_1   \; . \\
0 &  t_2 < t_1  \end{array} \right. \label{prop}
\end{equation}
            
We will now find an expansion for $\< \xi|{\e}_i|\xi\> $, the
non-unitary evolution $|\xi \> $ experiences through the region of
CTCs, by using the Born approximation. We will expand the complete
interaction diagrammatically and only look at terms which are first
order in $\l $.

We define $K(z_2,t_2;z_1,t_1)$ as the amplitude for all paths which
begin at $(z_1,t_1)$ and end at $(z_2,t_2)$. This includes all
scatterings and all numbers of windings through the identified
regions.  Expanding $K$ order by order in $\l$, we get
\be \label{kexp}
K = K_0 + \l K_1 + {\l}^2K_2 + ... \; ,
\ee where $K_i$ is the piece of the amplitude which is ith order in
the interaction.  \\ Using our propagators we will now find an
expression for $\< x'|\e |x\> $, which will be used to find $\< \xi
|\e |\xi \> $. Given our definition of $K$, we can expand $\<
x'|X^{\dagger}X|x\> $ in terms of propagators:
\begin{eqnarray} \label{XandK}
\< x'|X^{\dagger}X|x\> &=& \int dx_1 \, \< x'|X^{\dagger}|x_1\> \< x_1|X|x\> 
\; ,\nonumber \\
&=& \int dx_1 \, K^\ast (x_1,t_+;x',t_-)K(x_1,t_+;x,t_-) \; .
\end{eqnarray}
Arranging the terms order by order in $\l $,
\ba 
\< x'|X^{\dagger}X|x\> &=& \int dx_1 \left[ K_0 ^\ast(x_1,t_+;x',t_-)
K_0(x_1,t_+;x,t_-) \right. \nonumber \\
&&+ \; \l \left( K_0 ^\ast(x_1,t_+;x',t_-)K_1(x_1,t_-;x,t_-) \right.
\nonumber \\ &&+ \left. \left.  K_1 ^\ast(x_1,t_+;x',t_-)K_0(x_1,t_+;x,t_-)
\right) + o(\l ^2) \right] \; . \label{XX1}
\ea
$K_0$ is a free propagator, therefore
\be
\int dx_1 \, K_0 ^\ast(x_1,t_+;x',t_-)K_0(x_1,t_+;x,t_-) = 
\delta(x',x) \; .
\ee
If we define 
\be \label{Adef}
A(x,x') \equiv \int dx_1 \, K_0 ^\ast(x_1,t_+;x',t_-)K_1(x_1,t_+;x,t_-) \; ,
\ee
and throw away terms $o(\l ^2)$ and higher, then (\ref{XX1})  becomes
\be \label{XX2}
\< x'|X^{\dagger}X|x\> \approx \delta(x',x) + \l [A(x,x')+A^\ast(x',x)] \; .
\ee
Comparing (\ref{XX2}) to (\ref{Xexpansion2}), we find
\be
\< x'|\e |x\> = A(x,x')+A^\ast(x',x) \; .
\ee
We ignore terms $o(\l ^2)$ and higher, since we are using the Born
approximation.  We wish to investigate the non-unitary evolution in
equation (\ref{prob}), which is expressed by terms of the form $ \<
\xi|\e |\xi\> $, where $|\xi\> $ is a single-particle state.
\ba
\< \xi|\e |\xi\> &=&\int \! \! \int dx \, dx' \, \xi(x,t_-)[A(x,x')
+A^\ast(x',x)] \xi^\ast (x',t_-)\nonumber \\ &=& 2 \, {\rm Re} 
\left[ \< \xi|A |\xi\> \right] \label{Aamp}
\ea
$K_i$ includes every number of windings through the identified
regions.  Each $K_i$ can be expanded as a series, where each
subsequent term represents another winding through the time
machine.  For example, the expansion for $K_0$, illustrated in figure 3, is
\ba
K_0(x',t_+;x,t_-) = K_f(x',t_+;x,t_-) + \int_{-\frac{L}{2}} ^{\frac{L}{2}} dy 
K_f(x',t_+;y,t_-)K_f(y,t_+;x,t_-) \nonumber \\ + \int_{-\frac{L}{2}}
 ^{\frac{L}{2}} \int_{-\frac{L}{2}} ^{\frac{L}{2}} dy \,  dy' K_f(x',t_+;y,t_-)
K_f(y,t_+;y',t_-) K_f(y',t_+;x,t_-) + ...
\ea

\begin{figure}
\leavevmode
\centering
\epsfxsize=15cm 
\epsfbox{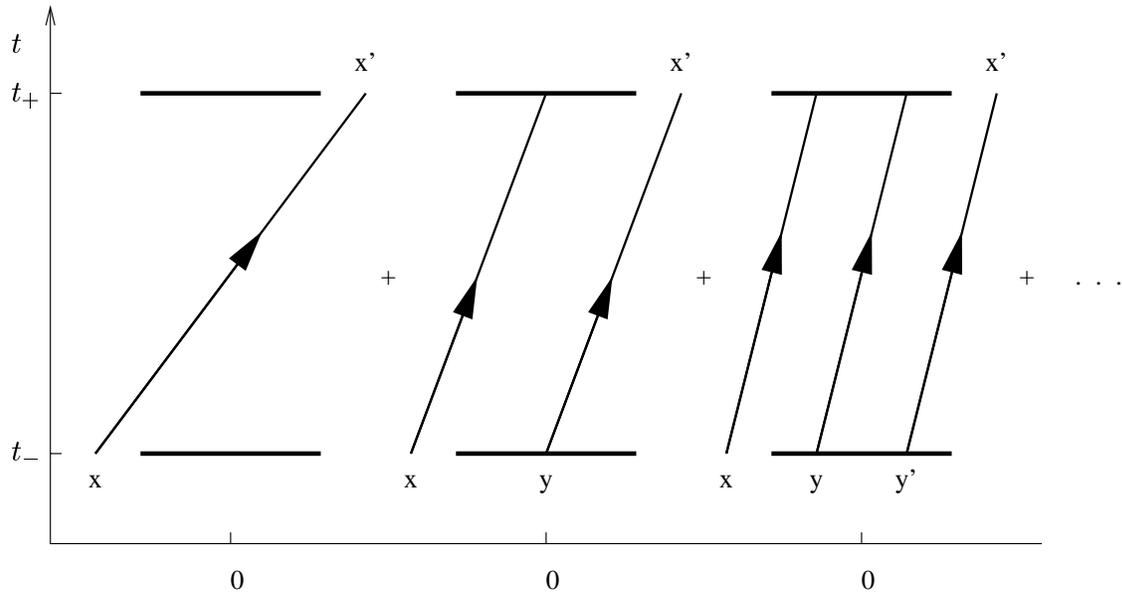}
\caption{The first three diagrams in the expansion of
$K_0(x',t_+;x,t_-)$ in terms of number of windings through the time
machine.  The figure includes diagrams with winding number 0, 1, and
2.
\label{fig3}}
\end{figure}

As we see in (\ref{Adef}), $A(x,x')$ can be expressed in terms of the
$K_i$'s.  Therefore, $A(x,x')$ is also the sum of an infinite number
of terms. We will initially simplify the calculation as Politzer does
\cite{politzer} by assuming $\hbar T >> mL^2$. For a given $K_i$,
each subsequent term has another winding through the time machine and
therefore another integral over $y$.  The assumption $\hbar T >> mL^2$
can be interpreted as choosing $T$ and $L$ such that the wave packet
spreads significantly compared to $L$ during the time $T$.  For each
additional winding through the identified regions, more of the wave
packet will spread outside $Y$ and not return to $t_-$ to wind through
again.  Thus, for each subsequent winding, less and less of the wave
packet will remain to contribute to the non-unitary evolution of the
next winding.  Thus any given term in the expansion is smaller than
the previous term by a factor of the small parameter
$\frac{mL^2}{\hbar T}$. We can therefore approximate each $K_i$ by the
first term in its expansion, i.e., the term with the lowest number of
windings.  With this assumption, the diagrammatic representation for
$A(x,x')$ is shown below in figure 4.

\begin{figure}
\leavevmode
\centering
\epsfysize=15cm 
\epsfbox{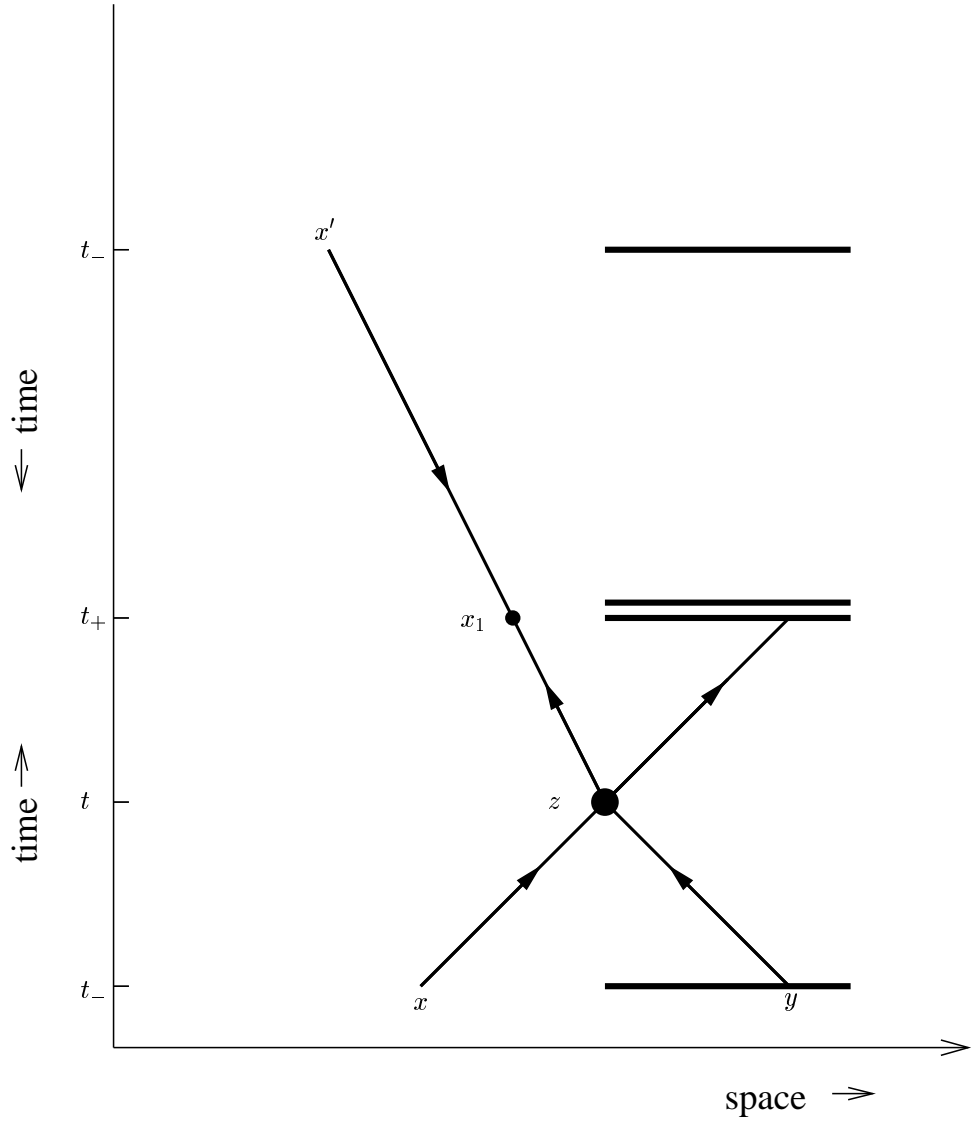}
\caption{The diagrammatic representation of $A(x,x')$: the $o(\l )$
term in the expansion of $\< x'|X^{\dagger}X|x\> $. Lines with
forward (backward) pointing arrows denote factors of $K_f$
($K_f^\ast$), the heavy horizontal lines denote the identified
surfaces, and the dot denotes a factor of $\l V_0$.
\label{fig4}}
\end{figure}

Since we are interested in estimating the dependence of $\<
\xi|A|\xi\> $ on the particle's spatial distance from the center of
$Y$, we will take $|\xi\> $ to be a stationary Gaussian wave packet
centered at $x=x_c$ with width $d$:
\be \label{gaussian}
\xi(x,t_-) = \pi^{-\frac{1}{4}}d^{-\frac{1}{2}}\exp \left[ \frac{-(x-x_c)^2}
{2d^2}\right] \; .
\ee
We are looking for the dependence of $\< \xi|A|\xi\> $ on $x_c$ for large
$x_c$.  Thus we assume
\ba
x_c &>>& L  \nonumber \; ,\\
x_c &>>& d \; ,\\
x_c &>>& \frac{\hbar T}{m} \nonumber \; .
\ea

An expression for $\< \xi|A|\xi\> $ is found from its diagrammatic
representation.  The full calculation is completed in the appendix,
and the dependence on $x_c$ extracted.  In essence, the integral is an
overlap between $|\xi\> $ and a particle caught in $R$.
Since we are using a contact potential, the overlap integral will fall off
exponentially as the Gaussian wave packet is moved farther from the
time machine. Thus, the dominant $x_c$ behavior is found to be
\be \label{Aande}
\< \xi|\e |\xi\> = 2 \,{\rm Re} \left[\< \xi|A|\xi\> \right] \propto \exp 
\left[ \frac{-x_c ^2}{d^2} \left(\frac{2R^2 +1}{2(R^2 +1)} \right) 
\right] \; , 
\ee
where
\[ R= \frac{md^2}{\hbar T} \; . \]
Due to the contact potential, the non-unitary evolution of $ | \xi \>
$ is proportional to the overlap between the wave packet and the
region $Y$. We interpret the result above to mean that the effect
falls off exponentially with the spatial distance between the center
of the wave packet and the center of $Y$.  Therefore, we expect $\<
\xi|\e |\xi\> \approx 0$ unless there is a significant overlap between
the wave packet and the region $Y$.  If $d < L$, the width of the
wave packet is smaller than that of $Y$, and we expect a significant
overlap only when the center of the wave packet is inside $R$, i.e.
$\< \xi|\e |\xi\> \approx 0$ unless $|x_c| < \frac{L}{2}$ for some
time between $t=t_-$ and $t=t_+$.  If $L < d$, the width of the
particle is larger than that of the time machine, and we expect that
$\< \xi|\e |\xi\> $ may be significant even if $|x_c| > \frac{L}{2}$.
In this case, we expect the overlap between the wave packet and the
region $(-\frac{L}{2}, \frac{L}{2})$ to be non-negligible as long as
the wave packet is positioned such that the center of $Y$ lies a
distance less than $d$ from the center of the wave packet.

Now we generalize our result to include the case when the minimal
winding approximation ($\hbar T >> mL^2$) does not hold.  In this
case, there are an infinite number of $o(\l )$ terms in our expansion
for $\< \xi|\e |\xi\> $.  These terms represent all possible numbers
of windings through the identification, and each term can be
represented by a diagram.  The summation of these terms is
non-trivial.  However, we are only interested in the dependence of $\<
\xi|\e |\xi\> $ on $x_c$ for large $x_c$. In each term of the
expansion $x_c$ will appear only in $\xi(x)$ and $\xi^\ast(x)$. Thus,
the dependence of each term on $x_c$ will only come from $\xi(x)$,
$\xi^\ast(x)$, and the propagators which connect the external wave
packets to the time machine. The propagators which connect one
identified region to the other, or one of the identified regions to
the interaction point will not contribute the to the $x_c$ dependence
of the term.  We will therefore organize the infinite set of diagrams
into 3 distinct classes.  Each diagram has two external propagators
which connect $\xi(x)$ and $\xi^\ast(x)$ to the identified
regions. Since we are limiting our expansion to $o(\l )$ each term
will be represented by a diagram with only one interaction point. The
three classes will be the sets of diagrams where both, one, or neither
of the external propagators connect to the interaction point.  We will
call these classes A,B, and C respectively. The lowest winding
diagrams are shown below in figure 5, grouped by class.

\begin{figure}
\leavevmode
\centering
\epsfysize=15cm 
\epsfbox{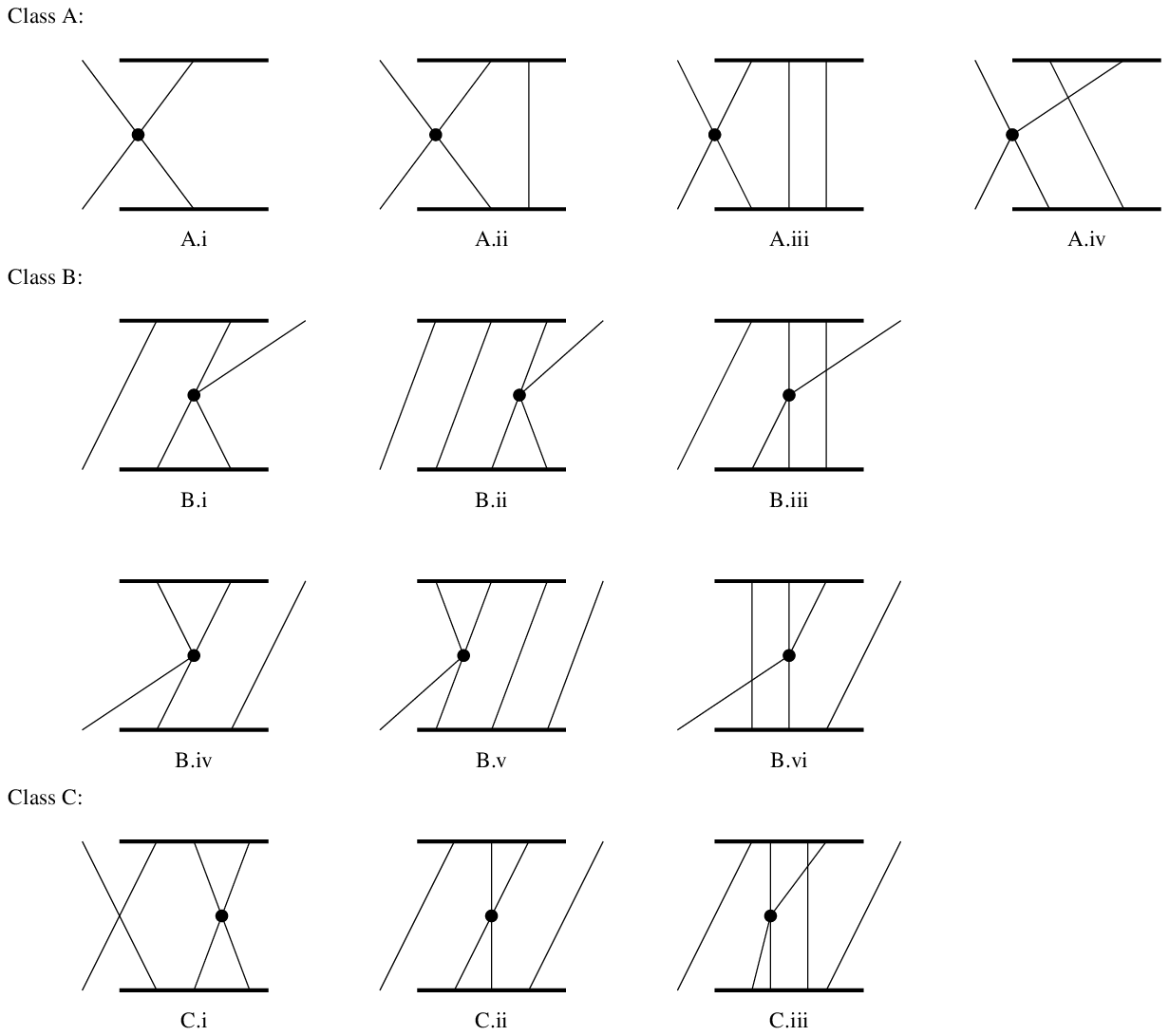}
\caption{The diagrams of $o(\l )$ with the lowest numbers of windings
through the identification in the expansion of $\< \xi|\e |\xi\> $.
These diagrams are split into three classes: A, B, and C, which
include diagrams with two, one, or zero external propagators connected
to the interaction point respectively.  The interaction point in each
diagram is denoted by a black dot. The intersection of two propagator
lines without a dot does not represent an interaction. The ${K_f}^\ast
$ term is suppressed in each diagram.
\label{fig5}}
\end{figure}

As discussed above, all diagrams in a given class will have the same
dependence on $x_c$. The $x_c$ dependence of diagram (A.i) was
calculated in the appendix, and is shown in (\ref{Aande}).  Therefore,
all diagrams in class A behave like $\exp \left(-\frac{x_c ^2}
{d^2}\right)$ for large $x_c$.  Similar calculations were done for
diagrams (B.i) and (C.i).  Both were found to behave like $\exp
\left(-\frac{x_c ^2}{d^2}\right)$ for large $x_c$, and so all
$o(\l )$ diagrams are dominated by a factor of $\exp \left(-\frac{x_c
^2}{d^2}\right)$ for large $x_c$. Therefore, all conclusions made
about the non-unitary evolution for the minimal winding case ($\hbar T
>> mL^2$) also hold when all windings are included.  Since the
non-unitary component of the evolution drops off exponentially with
the spatial distance between the wave packet and $Y$, the non-unitary
component of the evolution will be non-negligible only if the particle
passes through $R$, the region of spacetime between the identified
surfaces.

\section{The cross-section of a scattering experiment with a region of
CTCs in the experiment's future}

	In section 2 we found that a region of CTCs could alter the
probability assigned to a given history, even if the region of CTCs
was in the future of the last alternative in the history. Equation
(\ref{prob}) expresses how the CTCs alter the probability of any
transition occurring before them. In section 3 we found that for
histories which involve wave packets $\< \xi|\e |\xi\> \approx 0$,
unless there is significant overlap between the wave packet and the
region $R$.  For particles smaller than the identified spatial region,
this means that the particle must pass through $R$: the region of
spacetime between the two identified regions.  We will now examine a
scattering experiment, which provides us with states with well known
trajectories.  By examining which pre- and post-scattering states pass
through $R$, we will calculate the effect which the CTCs have on the
total cross-section.

The spacetime chosen here will be the same as that of section 3 except
that we will be working in 3+1 dimensions.  We choose a ball of
diameter $L$, centered at a point $p$, and label this region $Y^3$.
An identification is made between $Y^3$ at $t=t_-$ and $Y^3$ at
$t=t_+$. We define $R^3 \equiv Y^3 \otimes (t_-,t_+)$.  We assume our
initial density matrix $\rho$ has been prepared at $t=0$, and that it
describes a system of two particles, each with mass $m$.  The
particles are aimed at each other, each with speed $v$ as measured
from the lab frame.  Therefore, $ \vec{P} _{tot}=0$, and the
scattering will be spherically symmetric. As we are only interested in
estimating this effect, we will assume a semi-classical picture for
the particles: the position and momentum of each particle are
reasonably well known.  At time $t_2$, after the scattering, the
system is in the state $|\psi_i\> $.  Examination of the initial and
final states allows us to deduce the location of the scattering
center: $(x_s,t_s)$.  The spatial distance between the scattering
center and point $p$ is $D$, while the time difference between the
scattering event and the first identified region, $t_- - t_s$, we will
call $\cal{T}$.  This setup is shown below in figure 6.

\begin{figure}
\leavevmode
\centering
\epsfysize=15cm 
\epsfbox{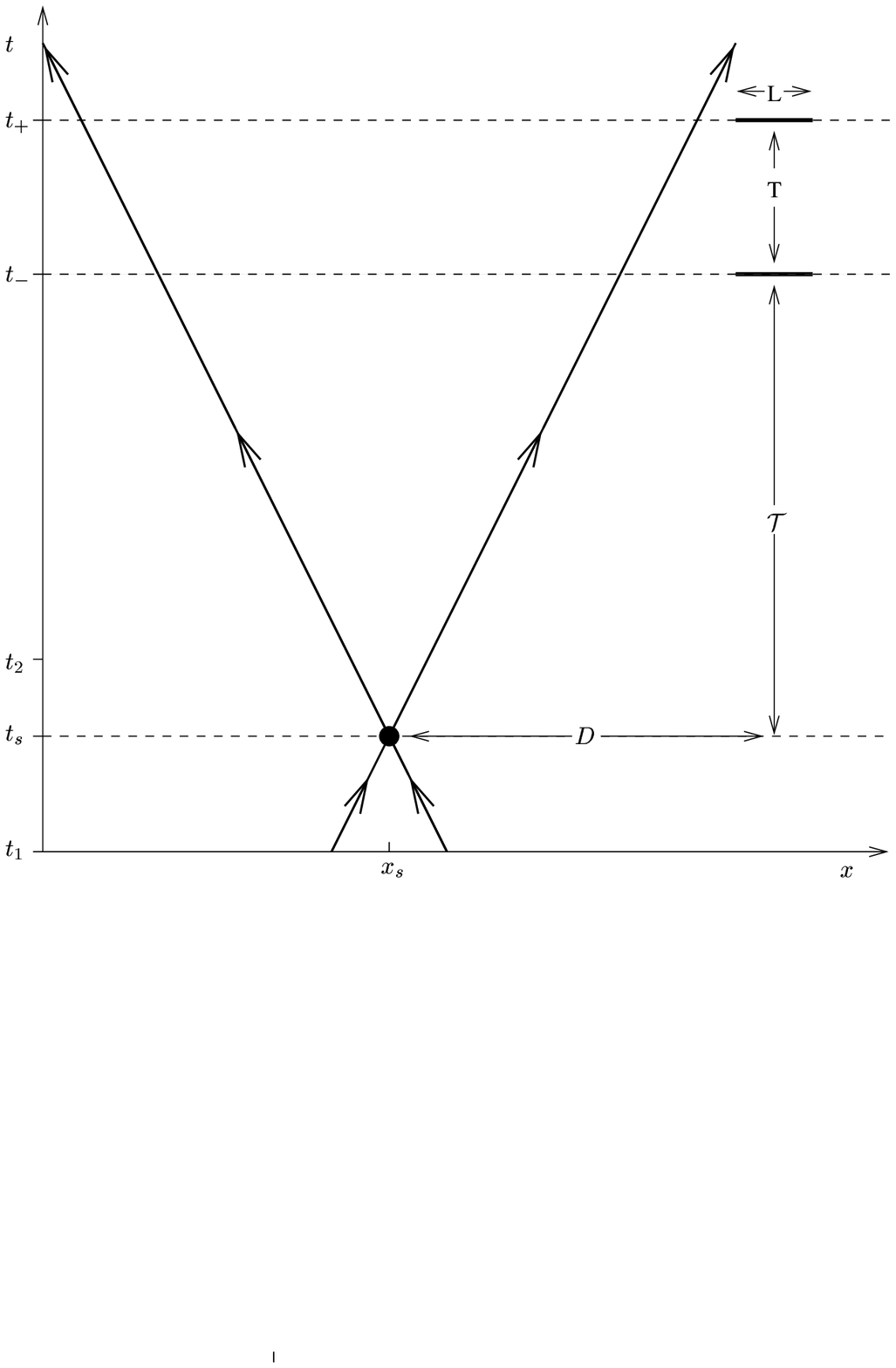}
\caption{The scattering event and its location with respect to the Politzer
identification.  The event takes place in 3+1 dimensions, so two spatial
dimensions are suppressed.
\label{fig6}}
\end{figure}

Equation (\ref{prob}) shows that the transitional probability
$p(\psi_i)$ is equal to the product of $\amp $, the transitional
probability if no CTCs exist, and some factor which represents the
non-unitary evolution of both the pre- and post-scattering states:
\be \label{shortprob}
p(\psi_i) = g(\psi_i,\phi) {\left|\< \psi_i|\phi\> \right|}^2 \; ,
\ee
where
\[ g(\psi_i,\phi) \equiv \frac{1+\l ( \< \psi_{i1}|{\e}_1
|\psi_{i1}\> +  \< \psi_{i2}|{\e}_2 |\psi_{i2}\> ) } {1+\l ( \< \phi_1|{\e}_1
|\phi_1\> +  \< \phi_2|{\e}_2 |\phi_2\> ) } \; . \]
From scattering theory we know that the differential cross-section,$\dcs $,
is directly proportional to the transitional probability:
\be
\dcs = Kp(\psi_i) \label{diff} \; .
\ee
The same relationship holds true if the CTCs do not exist.  Let
$\dcs _0$ be the differential cross-section if CTCs do not exist, 
\be
\dcs _0 = K\namp \; .
\ee
Where we will suppress the index $i$ from now on.  Combining
(\ref{shortprob}) and (\ref{diff}) we see that our differential-cross
section is related to the standard (no CTC) differential cross-section
by the same factor $\ng() $ which relates $p(\psi)$ and $\namp $.

We integrate over $d\o $ to get the total cross-section.
\ba
\s &=& \int_{4\pi} \dcs d\o \; ,\nonumber \\
   &=& \int_{4\pi} \dcs _0 \ng() \, d\o \; ,\nonumber \\
   &=& \int_{4\pi} \dcs _0 \left[ \frac{1+\l ( \< \psi_{1}|{\e}_1
|\psi_{1}\> +  \< \psi_{2}|{\e}_2 |\psi_{2}\> ) } {1+\l ( \< \phi_1|{\e}_1
|\phi_1\> +  \< \phi_2|{\e}_2 |\phi_2\> ) } \right] d\o \; .
\ea
The initial states, $|\phi_1\> $ and $|\phi_2\> $, are independent of the 
scattering angle, while the post-scattering states $|{\psi}_1\> $ and 
$|{\psi}_2\> $, depend on the scattering angle.  Thus, only the 
post-scattering states will vary as we integrate over $ d\o $.
\ba 
\s &=& \frac{1}{1+\l(\< \phi_1|{\e}_1|\phi_1\> +\< \phi_2|{\e}_2 |\phi_2\> )}
\int_{4\pi} \dcs _0 \left[ 1+\l ( \< \psi_{1}|{\e}_1|\psi_{1}\>  
+ \< \psi_{2}|{\e}_2 |\psi_{2}\> ) \right] \, d\o ,  \nonumber \\
&=& \frac{1}{1+\l(\< \phi_1|{\e}_1|\phi_1\> +\< \phi_2|{\e}_2 |\phi_2\> )}
\left[ \s _0  + \l \int_{4\pi} \dcs _0 \< \psi_{1}|{\e}_1|
\psi_{1}\> \, d\o  \nonumber \right. \\ 
&& \qquad \qquad  \qquad \qquad \qquad  \qquad \qquad \; \; \; + \left. \l 
\int_{4\pi} \dcs _0 
\< \psi_{2}|{\e}_2 |\psi_{2}\> \, d\o \right] \; . \label{tcs1}
\ea
Where $\s _0$ is the total cross-section if CTCs do not exist.

As discussed in section 3, for a given wave packet $|\xi\> $, $\<
\xi|\e |\xi\> \approx 0$ unless a significant portion of the wave
packet passes through $R^3$. This will only occur for certain values
of the particle's momentum. First, only a limited range of particle
speeds will enable the particle to reach $Y^3$ between $t=t_-$ and
$t=t_+$.  The trajectory of the particle represented by $|\xi\> $ must
lie inside the shaded region shown in figure 7, in order for $\<
\xi|\e _j |\xi\> $ to be non-negligible.

\begin{figure}
\leavevmode
\centering
\epsfysize=15cm 
\epsfbox{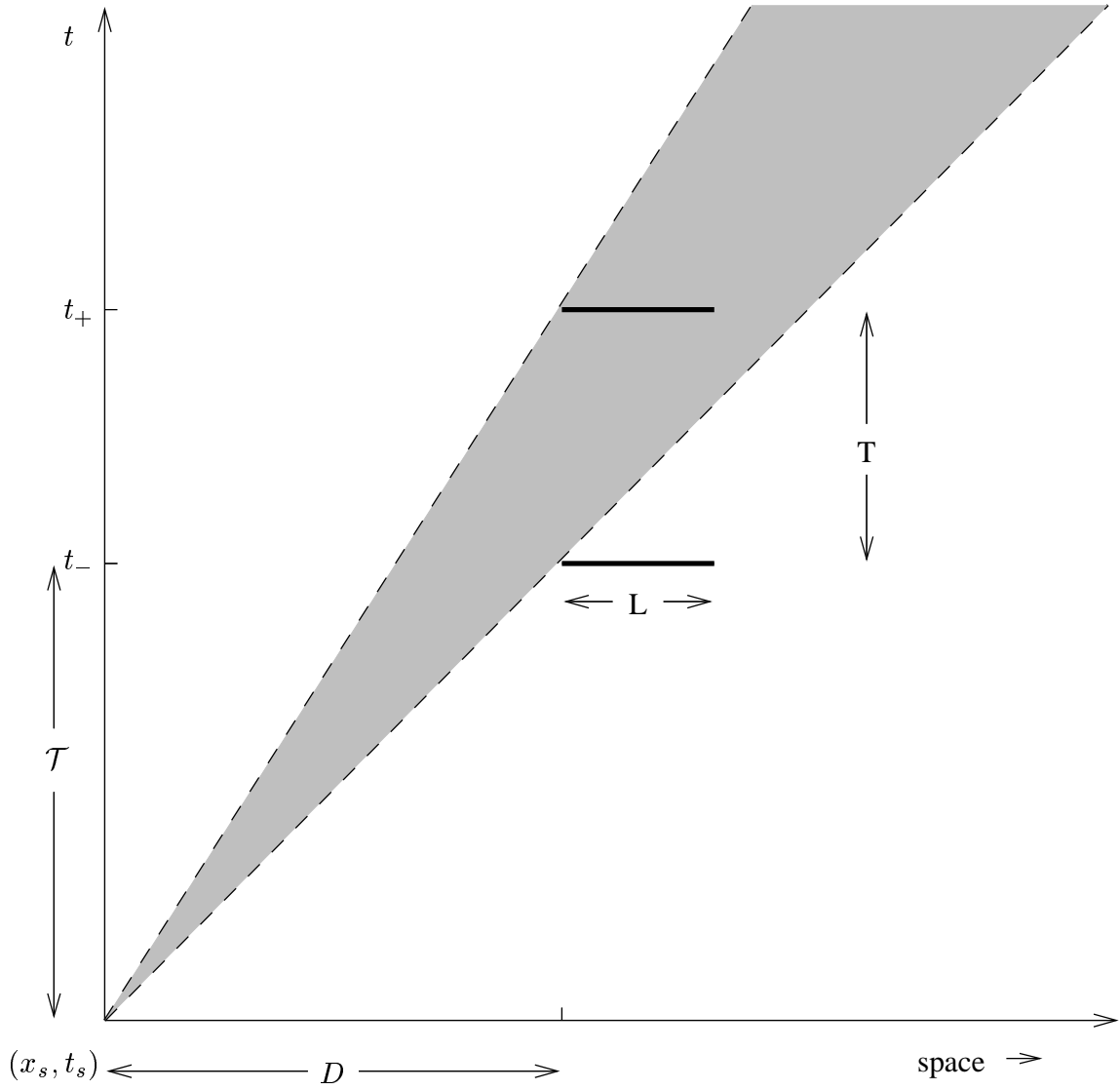}
\caption{An illustration of the range of speeds for a post-scattering
particle which lead to a possible non-unitary effect. The particle's
trajectory must pass through the shaded area in order for the particle
to have any chance of passing inside the time machine.
\label{fig7}}
\end{figure}

This leads to the following range for $v_i$, the speed of the particle in
question:
\be \label{velrange}
\frac{D}{\cal{T}} < v_i < \frac{D}{{\cal{T}} +T} \; .
\ee

Due to our chosen initial condition $v_{10}=v_{20}=v_{1f}=v_{2f}=v$.
For now, we assume that $v$ falls within this range.  Even if this
condition is met, the particle must be traveling towards $Y^3$ in
order to pass inside it. This will have different consequences for the
pre- and post-scattering particles, since we will integrate over all
final angles that the the post-scattering particles scatter into. The
trajectories of the initial particles are fixed once the initial state
is chosen.  Therefore, if neither of the initial particles is aimed at
$Y^3$ then $\< \phi_j|\e |\phi_j\> \approx 0 \, , \; (j=1,2)$.  For
now, we will assume this to be the case and examine the effect on the
total cross-section due to the post-scattering particles only.
With this assumption equation (\ref{tcs1}) becomes
\be \label{tcs2}
\s = \s _0 +\l \int_{4\pi} \dcs _0 \< \psi_{1}|{\e}_1|\psi_{1}\> \, d\o   
+\l \int_{4\pi} \dcs _0 \< \psi_{2}|{\e}_2 |\psi_{2}\>  \, d\o \; .
\ee
We define
\be \label{changetcs}
(\Delta\s )_j \equiv \l \int_{4\pi}\dcs _0\< \psi_{j}|{\e}_j|\psi_{j}\> 
\, d\o \; , 
\ee
where $j=(1,2)$, as the the change in the total cross-section due to
particle $j$. The post-scattering particles must also be aimed at the
time machine in order for $\< \psi_{ij}|\e_j|\psi_{ij}\> $ to be
significant.  To find the total cross-section, we will integrate over
$d\theta \, d\phi $, which represents all angles which one of the
particles could scatter into. The other particle will, in this case,
simply scatter off the scattering center with an angle $\pi$ relative
to the first particle. For a given particle, only a fraction of
post-scattering angles will direct the particle towards $Y^3$.  This
fraction, $F$, is the fraction of the sphere subtended by the time
machine as viewed from the scattering center.  $F$ is related to $\o
_{tm}$, the solid angle subtended by $Y^3$ as viewed from the
scattering center. $\o _{tm}$ is illustrated in figure 8.  $F$ is
given in terms of $\o _{tm}$ by

\begin{figure}
\leavevmode
\centering
\epsfxsize=15cm 
\epsfbox{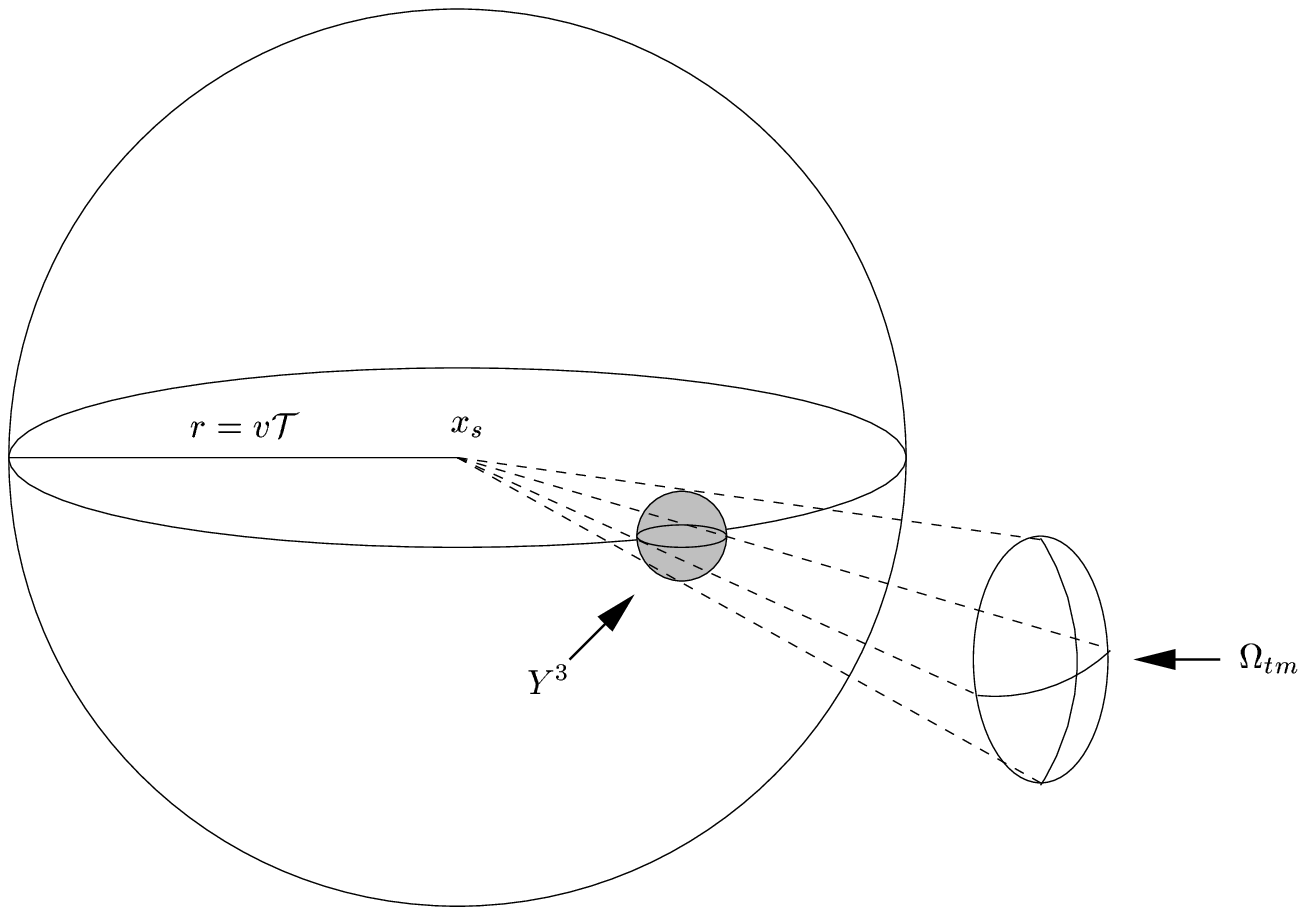}
\caption{The solid angle subtended by $Y^3$ as viewed from the
scattering center is $\o _{tm}$.  Only post-scattering states
aimed towards $Y^3$ will experience any non-unitary evolution.
\label{fig8}}
\end{figure}

\be \label{solidangle}
F = \frac{\o _{tm}}{\o _{tot}} = \frac{\pi {\left( \frac{L}{2} \right)}^2}
{4\pi D^2} = \frac{L^2}{16D^2} \; .
\ee
When we integrate equation (\ref{changetcs}) over $d\theta \, d\phi$, 
$\< \psi_{j}|{\e}_j|\psi_{j}\> \approx 0$ unless the element of the sphere 
which we are integrating over is contained in the fraction of the sphere
subtended by $Y^3$.
\be \label{changetcs2}
(\Delta\s )_j = \l \int_{\o_{tm}}\dcs _0\< \psi_{j}|{\e}_j|\psi_{j}\> 
\, d\theta \, d\phi \; , \; j=(1,2) \; .
\ee
However, $\< \psi_{j}|{\e}_j|\psi_{j}\> $ is not constant over the fraction
of the sphere subtended by $Y^3 \, $. As shown in the appendix, 
\be \label{nonunitary}
\< \psi_{j}|{\e}_j|\psi_{j}\> \approx \frac{-2V_0}{\pi} {\left( \frac{m^3}
{\hbar ^5 T^3} \right) }^\half h(d,L)\, {T'}^3 \; , \; h(d,L)= 
\left\{ \begin{array}{ll}1 & d<<L \\ \frac{L}{ed} & L >> d \end{array} 
\right. \, .
\ee
Here $T'$ is the time spent by the particle in $Y^3$. Let us define
$S$ to be the length of the particle's path inside $Y^3$, which $T'$
is clearly dependent on.  Assuming the particle does not wind through
the identification \footnote{This assumption is reasonable since we
are interested in an order of magnitude estimate for the deviation in
the total cross-section.  An extra winding will vary the result by one
order of magnitude at most.}, $S$ will vary with the $\theta$ and
$\phi$ which the particle is scattered into. If we align our $\theta =
0$ axis along the line connecting the scattering center and point $p$,
we have azimuthal symmetry, and find
\[ S = \left( L^2 - 4D^2 {\sin ^2\theta} \right)^\half \]
and
\be \label{time}
T' = \frac{ \left( L^2 - 4D^2 \sin ^2\theta \right) ^\half }{v} \; .
\ee
Substituting (\ref{time}) and (\ref{nonunitary}) into (\ref{changetcs2}) 
we find
\be \label{changetcs3}
(\Delta\s )_j = \frac{-2\l V_0 h(d,l)}{\pi v^3}{\left( \frac{m^3}{\hbar ^5 
T^3} \right)}^\frac{1}{2}  \int_{\o_{tm}} \dcs _0 \left( L^2 - 4D^2 {\sin 
^2\theta} \right)^\frac{3}{2}  \sin \theta \, d\theta \, d\phi \; .
\ee
Due to our choice of spherically symmetric scattering event, we know
\be \label{ss}
\dcs _0 = \frac{\s _0}{4\pi} \; .
\ee
However, if the scattering is not spherically symmetric, then we simply
substitute the correct expression in for $ \dcs _0 $.

Combining equations (\ref{changetcs2}), (\ref{nonunitary}), (\ref{time}),
(\ref{solidangle}), and (\ref{ss}) we find
\be \label{changetcs4}
(\Delta\s )_j = \s _0 \left\{ \frac{-2\l  V_0 h(d,L) }{ \pi v^3} {\left( 
\frac{m^3}{\hbar ^5 T^3} \right)}^\frac{1}{2}  \frac{L^5}{40 D^2}
\left[ 1 - \left( \frac{L^2}{L^2 + 4D^2} \right)^\frac{5}{2} \right] \right\}
\; .
\ee
The effect will be the same for each of the two particles, so $(\Delta\s )_1
=(\Delta\s )_2 $.  Therefore
\be \label{dev}
\s = \s _0 \left\{ 1 -  \frac{ \l L^5 V_0}{10 \pi D^2 v^3} {\left( 
\frac{m^3}{\hbar ^5 T^3} \right)} ^\frac{1}{2} h(d,L) \left[ 1 - \left( 
\frac{L^2}{L^2 + 4D^2} \right) ^\frac{5}{2} \right] \right\} \; .
\ee

We have arrived at an expression for the deviation in the total cross-
section of a scattering experiment with a region of CTCs in its
future.  This expression assumes that neither pre-scattering particle is
aimed towards $Y^3$ and that the velocity of the particles is within
the range specified in (\ref{velrange}).

	We have shown that with Hartle's prescription for calculating
probabilities in spacetimes with CTCs the total cross-section of a
scattering experiment before the region of CTCs will deviate from the
standard (no CTCs) value as shown in (\ref{dev}). The question now
arises: will this effect be large enough to be measured in a typical
scattering experiment?  We will now give an estimate of the size of
this effect for a few cases.  If topological or geometric defects
which cause CTCs are allowed by the laws of physics we expect that the
scale for naturally occurring defects will be the Planck scale, for
both time and space \cite{wheeler,halliwell}.  We will model a defect
of this kind using a Planck-scale Politzer time machine.  We plug in
reasonable values for electron-electron scattering into (\ref{dev})
and take a Planck-scale time machine a Planck length away from the
scattering center.  If we assume that the condition on the velocity of
the electrons given in (\ref{velrange}) is met, then the deviation in
the total cross-section is approximately $1$ part in $10^{80}$.  If
the Planck-scale time machine is $1$ meter from the scattering center,
the deviation drops to $1$ part in $10^{150}$. If we include the
possibility of man-made time machines, which could be of a size
traversable by humans~\cite{wormhole,Gott}, then the deviation becomes
larger. For a identified region 1 meter from the scattering center
with $L=1 \,$m and $T=1 \, $s, the deviation is $1$ part in
$10^{15}$. For the purpose of comparison to typical experimental error
of total cross-sections we will use the measurement of the anomalous
magnetic moment of the electron, which has an accuracy of $1$ part in
$10^{11}$~\cite{pdg}.  We will use this result as a upper bound on the
sensitivity of total cross-section measurements. Therefore, the
deviation we have calculated in the total-cross section due to a
Politzer time machine in our future is negligible compared to the
experimental error associated with the measurement. Therefore, even if
the time machine is located such that (\ref{velrange}) is satisfied it
would be impossible to detect a single Planck-scale or human-scale
time machine in our future from its effect on the total cross-section
of a scattering event.

If multiple time machines exist, then for a given scattering event the
only ones which will contribute to a deviation in the total
cross-section will be those located such that the velocity of the
particles satisfies (\ref{velrange}). We have found that the effect is
additive for multiple time machines which satisfy this
condition. Again, we assume that neither of the pre-scattering
particles are aimed at any of the time machines which contribute to
the deviation. The total cross section for a specific scattering event
can be computed by selecting only the time machines for which the
velocity satisfies (\ref{velrange}), calculating the deviation in the
total cross-section for each time machine using (\ref{dev}) and adding
the deviations for all the time machines selected. Therefore, the more
time machines we have in our future, the greater the chance of
observing a deviation in the total cross-section of a scattering event
in our present, as long as neither pre-scattering particle is aimed at
a time machine.  For a given event, if there are more time machines
then there is a greater chance that enough time machines will be at
correct locations in spacetime to satisfy (\ref{velrange}) and cause a
measurable deviation.

Because of the additive effect it is possible for a distribution of
time machines throughout spacetime to cause a deviation in the total
cross-section of a scattering event which would be measurable.  In
general, we cannot place limits on size, number or proximity of time
machines, because for any given values for these quantities, it would
be possible to design a configuration of time machines such that
(\ref{velrange}) was not satisfied for a given set of scattering
events.  In other words, the time machines could be distributed such
that the particles involved in the scattering events would never pass
inside any of the time machines, and therefore no deviation in the
total cross-section would occur. Also, these results are only valid if
neither pre-scattering particle is aimed at any of the contributing
time machines .  What is possible, however, is to rule out specific
spacetime distributions of time machines in light of specific
scattering experiments.  For example, if we take an electron-electron
scattering experiment for which the measured total cross-section
agrees with the standard predicted result (no CTCs) within the
experimental error, then we can rule out a spacetime which has either
Planck-scale or human-scale time machines everywhere in spacetime
except for the region of spacetime which surrounds the trajectory of
the pre-scattering particles. When we calculate the deviation in the
total cross-section due this distribution using (\ref{tcs1}), we find
that the term in the numerator, which comes from integrating the
non-unitary effect over all the post-scattering states is infinite.
Due to lack of time machines in the path of the initial particle, the
term in the denominator, which represents the non-unitary effect of
the pre-scattering particles, is simply $1$.  Therefore, Hartle's
prescription predicts an infinite deviation in the total cross section
in this case, which is clearly absurd, so such spacetimes are ruled
out.

The same result ensues if instead of the spacetime discussed above we
have an expanding shell of time machines: at any given moment in time
there is a spherical distribution of time machines around the
scattering center at the same radius as the post-scattering particles.
Again, the region surrounding the trajectory of the pre-scattering
particles is empty. This distribution of time machines has an
identical effect as the previous distribution, since this distribution
is simply the subset of time machines in the previous example for
which the particles' velocity satisfies (\ref{velrange}). It is also
possible to imagine finite versions of these examples which would
cause measurable deviations in the total cross-section. For either of
the examples above, a finite distribution of time machines would lead
to a measurable deviation in the total cross-section section of a
given experiment if the radius of the distribution is large
enough. For Planck-sized time machines, the radius of the distribution
must be approximately $10^{16}$\,light years which is extremely large.
However, a distribution of human-scale time machines (with the
dimensions discussed above) centered on the the scattering event need
only have a radius of 10\,km in order to cause a measurable deviation in
the total cross section.  If the distribution is not centered on the
event, or is at a distance from the scattering event, the effect is
diminished, and therefore we need a larger distribution in order for
the deviation to be measurable. If Hartle's prescription for applying
generalized quantum mechanics to spacetimes with CTCs is correct, then
spacetimes with any of these distributions of CTCs must be
disallowed. For reasonably sized time machines, one needs a large
number of time machines in close proximity to the scattering center in
order for a measurable deviation in the total cross-section to occur.
Clearly, there are a number of realistic distributions of Politzer
time machines which would lead to deviations in the total cross
section which are minute compared to the experimental error of any
given scattering experiment.  If the time machines are small, far
away, or sparsely distributed, then it is very possible that a region
of CTC's could go completely unnoticed.

	Some authors~\cite{wheeler,halliwell} have asserted that
quantum gravity predicts that at the Planck scale, spacetime is filled
with CTCs.  One possibility is that spacetime can be viewed as a
quantum foam, with Planck-scale CTCs throughout spacetime.  Another
hypothesis is that CTCs will become prevalent during the big bang and
the big crunch, due to quantum gravitational effects which are large
at these times.  We can model these possibilities simply with
Planck-scale Politzer time machines which pervade spacetime. In any
of these cases one might expect a significant deviation in the total
cross-section of a scattering experiment due to the dense distribution
of time machines.  However, as we will show, this is not the
case. Each of these cases involves an isotropic distribution of
time machines: all pre- and post-scattering particles will pass
through the same distribution of time machine. We will now use
(\ref{tcs1}) to examine the total cross section of a scattering event.
We will assume that $\< \xi_i|\e _i| \xi_i\> $ is the total non-unitary
effect of all time machines which affected the particle $|\xi_i\>
$. ($|\xi_i\> $ is generic and could be any of the pre- or
post-scattering particles.) $\< \xi_i|\e _i|\xi_i\> $ will be identical
for all states, including all pre-scattering states, and all possible
post-scattering states, since the distribution of time machines
is isotropic.  When we calculate the total cross-section using
(\ref{tcs1}), the integrals over $d\theta \, d\phi$ give factors of
$4\pi$, which leads to the result:
\be
\s ' = \s
\ee
Thus, when the distribution of time machines is isotropic, there
is a cancellation effect, and the predicted total cross-section agrees
with result expected if no CTCs existed.  We see that this is a
limiting case as $\o _{ctc }$ approaches $4\pi$.  Therefore, we also
expect the effect to be small if the distribution is nearly isotropic,
or if only a small fraction of the solid angle has a different
distribution of CTCs.  In this case we would expect the effect caused
by the initial and final particles to be nearly identical as long as
the initial particles are not aimed at the small fraction of solid
angle with a different distribution of time machines from the rest.
\section{Acknowledgements} \nonumber
The author would like to thank A.~Anderson, D.~Craig, D.~Deutch,
C.~Fewster, J.~Friedman, S.~Hawking, S.~Kolitch, D.~Politzer, and
J.~Simon for their helpful comments and conversations. Special thanks
are given to H.~A.~C.~Chamblin, S.~Ross, and J.~S.~Whelan for their
time and support.  Finally, the author would like to thank his
advisor, J.~B.~Hartle, for without his help and patience this paper
would never have been finished.
\appendix
\section{Appendix}
We will be using the same notation for position coordinates $x,y,$ and $z$ as
in section 3.  Combining (\ref{Adef}) and (\ref{Aamp}) we have
\be
\< \xi|A |\xi\> =\int \! \!\int dx \, dx' \xi(x,t_-) \left[\int dx_1 K_0 ^\ast
(x_1,t_+;x',t_-)K_1(x_1,t_+;x,t_-) \right] \xi^\ast(x',t_-) \; .
\ee
where $K_0$ and $K_1$ are the amplitudes which are 0th and 1st order
in the interaction respectively. As discussed previously, if we take
the minimal winding approximation ($mL^2 << \hbar T$), $\< \xi|A|\xi\>
$ can be represented by the diagram shown in figure 4.  This leads to
the following expression for $\< \xi|A|\xi\> $: \ba \< \xi|A|\xi\> &=&
-\frac{i}{\hbar} \int \! \! \int dx \, dx' \int_{t_-}^{t_+}dt \int \!
 \! \int dz \, dx_1 \yint dy \, K_f(z,t;x,t_-)K_f(z,t;y,t_-) \nonumber \\
&\times & V_0 K_f(y,t_+;z,t) K_f(x_1,t_+;z,t)K_f^\ast(x_1,t_+;x',t_-) 
\xi(x,t_-) \xi^\ast(x',t_-) \; . \label{bigamp}
\ea
Now,
\ba
&&\int  dx K_f(z,t;x,t_-) \xi(x,t_-) = \xi(z,t) \; ,\nonumber \\
&&\int  dx_1 K_f(x_1,t_+;z,t)K_f^\ast(x_1,t_+;x',t_-) = K_f^\ast(z,t;x',t_-) 
\; ,\nonumber \\
&&\int  dx' K_f^\ast(z,t;x',t_-) \xi^\ast(x',t_-) = \xi^\ast(z,t) 
\; .\label{stuff}
\ea
Combining (\ref{bigamp}) and (\ref{stuff}),
\be \label{smallamp}
\< \xi|A|\xi\> = -\frac{iV_0}{\hbar} \int_{t_-}^{t_+}dt \int dz \yint dy
\, K_f(z,t;y,t_-) K_f(y,t_+;z,t) {\left|\xi(z,t)\right|}^2 \; .
\ee
\\
We chose $\xi(x,t_-)$ to be a stationary Gaussian 
wave packet of width $d$ centered at $x=x_c$ in (\ref{gaussian}).  This state 
will spread with time:
\be \label{gaussian2}
{\left|\xi(z,t)\right|}^2 = \frac{1}{d(t)\sqrt{\pi}}\; \exp 
\left[ \frac{-(z-x_c)^2}{{d(t)}^2}\right] \; ,
\ee
where
\[  d^2(t) = d^2 \left( 1 + \frac{\hbar^2 t^2}{m^2d^4} \right) \; .\]
Combining (\ref{smallamp}), (\ref{prop}), and (\ref{gaussian2}),
\be
\< \xi|A|\xi\> = -\frac{mV_0}{2\hbar^2 \sqrt{\pi ^3 T}}
\int_{t_-}^{t_+}dt \frac{1}{\tau^\half (t)d(t)} 
\int dz \yint dy \, \exp \left( \frac{im{(z-y)}^2}{2 \hbar \tau(t)} 
- \frac{{(z-x_c)}^2}{d^2(t)}\right) \; ,
\ee
where
\[ \tau(t) = \frac{t(t-T)}{T} \; .\]
Completing the square for $z$ and integrating,
\be \label{longint}
\< \xi|A|\xi\> = \frac{imV_0}{2\pi \hbar^2 \sqrt{T}}
\int_{t_-}^{t_+}dt {\left(\frac{1}{\tau(t) d^2(t) \s (t)}\right)}
^\half \,  \yint dy \, \exp \left[ -\G ^2(t){(y-x_c)}^2 \right] \; ,
\ee
where
\[ \s (t) = \frac{imd^2(t)-2\hbar \tau(t)}{2\hbar \tau d^2(t)} \qquad 
\G ^2(t) = \frac{im}{imd^2(t) -2\hbar \tau(t) } \; .\] 

As discussed in section 3, we are interested in the functional
dependence of $\< \xi|\e |\xi\> $ on $x_c$ for large $x_c$. We assume
$x_c >> L$, and $y$ is bounded by $\pm \frac{L}{2}$.  Thus $x_c >> y$
and we define a small parameter $\delta$ to be
\[ \delta \equiv \frac{y}{x_c} \; . \]
Expressing the exponential of our $y$ integral in terms of $\delta$, we get
\be
\exp \left[ -\G ^2(t){(y-x_c)}^2 \right] =  exp \left[ -\G ^2(t){x_c}^2
(\delta -1)^2 \right] \; .
\ee
Now, $ \delta << 1$, so we drop it from our expression and integrate.
\be
\< \xi|A|\xi\> = {\left(\frac{im}{2 \hbar^3 T} \right)}^\half 
\frac{V_0}{\pi} \int_{t_-}^{t_+}dt \, {\left[\G ^2(t)\right]}^\half  \, 
\exp \left[ -\G ^2(t){x_c}^2 \right] \; .
\ee

The dependence of $\< \xi|\e |\xi\> $ on $x_c$ for large $x_c$ can now
be found using the method of steepest descents to extract the large
$x_c$ behavior and taking the real part.  $x_c >> d$ so we define a
large parameter $\Delta$:
\[ \Delta \equiv \frac{{x_c}^2}{d^2} \; . \]
Expressing our integrand in terms of $\Delta$,
\be
\< \xi|A|\xi\> = {\left(\frac{im}{2 \hbar^3 T} \right)}^\half 
\frac{V_0}{\pi} \int_{t_-}^{t_+}dt \, {\left[\G ^2(t)\right]}^\half  \, 
\exp \left[ -\G ^2(t)d^2\Delta \right] \; .
\ee
Since $\Delta$ is large the greatest contribution to this integral
will occur when $ \G ^2(t)d^2 $ is a minimum.  We find the saddle
point by solving for the complex $u_0$ which satisfies
\be \label{saddle1}
\left. \frac{d\G ^2(u)}{du}\right|_{u=u_0} = 0 \; , \qquad  \G ^2(u) = 
\frac{im}{im\left(\frac{m^2d^4 + \hbar^2u^2}{m^2d^2}\right) - 2\hbar\left
(\frac{u(T-u)}{T}\right)} \; .
\ee
Solving (\ref{saddle1}) for $u_0$, we find
\be \label{saddle2}
u_0 = \frac{md^2T(2md^2 -i\hbar T)}{\hbar^2T^2 + 4m^2d^4} \; .
\ee
We find $ \G ^2(u_0)$ by substituting (\ref{saddle2}) into our expression
for $ \G ^2(u)$ in (\ref{saddle1}). We express $ \G ^2(u_0)$ in terms of 
a dimensionless parameter $R = \frac{md^2}{\hbar T}$,
\be
\G ^2(u_0) = \frac{1}{d^2}\left[ \frac{2R^2 +1}{2(R^2+1)} -\frac{iR}{2(R^2+1)}
\right]
\ee
The integral will be dominated by the integrand at $u = u_0$, so
\be
\< \xi|A|\xi\> \approx {\left(\frac{im}{2 \hbar^3 T} \right)}^\half 
\frac{V_0}{\pi } {\left[\G ^2(u_0)\right]}^\half  \, \exp \left[ -\G ^2(u_0)
d^2\Delta \right] \; .
\ee
$\< \xi|\e |\xi\> $ is proportional to $ {Re} \left[ \< \xi|A |\xi\>
\right] $ as shown in (\ref{Aamp}).  When we take the real part of $\<
\xi|A |\xi\> $ and extract the large $x_c$ dependence, the dominant
behavior is found to be due to the real part of the exponent:
\be
\exp \left\{ {\rm Re} \left[-\G ^2(u_0) d^2\Delta \right] \right\} = \exp 
\left[ - \left( \frac{2R^2 +1}{2(R^2+1)}\right)\Delta \right] \; .
\ee
Putting $\Delta $ in terms of $x_c$, we find the large $x_c$ behavior of
$\< \xi|\e |\xi\> $ to be
\be
\< \xi|\e |\xi\> \propto \exp \left[ \left( - \frac{2R^2 +1}{2(R^2+1)}\right)
\frac{{x_c}^2}{d^2} \right] \; .
\ee

In addition to finding the large $x_c$ behavior of $\< \xi|A|\xi\> $,
we are also interested in finding the exact expression for the
non-unitary evolution when the non-unitary evolution is significant.
As discussed in section 3, the non-unitary evolution is dependent on
the overlap between the the wave function and $Y$
between the times $t_-$ and $t_+$. We will no longer examine
stationary particles, but moving ones, whose trajectories are such
that the overlap between their wave packets and $Y$ are significant
between $t_-$ and $t_+$.  In general, it will be difficult to get a
simple analytic expression for the non-unitary evolution, so we will
examine the cases where the particle is much smaller or larger than
the identified region: $ L >> d$ and $L << d$ respectively. If $ L >>
d$ this corresponds to the particle passing through $R$, while if $L
<< d$, then the particle's trajectory must be such that the center of
$Y$ passes within $d$ of the wave packet's center between $t_-$ and
$t_+$.

We begin with (\ref{longint}) and make a change of variables, letting
$w=\G {(t)(y-x_c)}$.
\be
\< \xi|A |\xi\> = {\left( \frac{im}{2 \hbar^3 T} \right)}^\half 
\frac{V_0}{\pi } \int _0 ^T dt \int _{-\G (\frac{L}{2}+x_c)} ^{\G (\frac{L}{2}
-x_c)} dw \; \exp (-w^2) \; .
\ee
Now we will assume that the wave function's overlap with $Y$ will be
constant for the entire time that the overlap is significant.  Thus,
if $ L >> d$, then the width of the wave packet will stay much smaller
than $L$ for the entire time that the particle is inside $Y$, while if
$L << d$ the wave packet is very wide, and will not spread enough to
change the overlap inside a half- width significantly.  Thus, if $ L
>> d$, then the Gaussian integral will yield approximately a factor of
$\sqrt{\pi} $ when the particle is inside $R$ , while if $L << d$, the
integral yields approximately a factor of $ \frac{\sqrt{\pi} L}{ed}$
when the center of $Y^3$ is within $d$ of the center of the wave
packet. If we assume that the particle's motion relative to $Y^3$ will
cause the particle to pass through $R$ for a finite period of time,
which we will call $T'$, then the the integral over time will yield a
factor of $T'$.  Therefore (\ref{longint}) reduces to
\be \label{directhit1}
\< \xi|A |\xi\> \approx {\left( \frac{im}{2 \pi \hbar^3 T} \right)}^\half V_0 
\, T' \, h(d,L) \, ,
\ee
where
\[ h(d,l)= \left\{ \begin{array}{ll} 1 & d<<L \\ \frac{L}{ed} & L >> 
d \end{array} \right. \, . \]
Combining (\ref{directhit1}) and (\ref{Aamp}), we get
\be \label{directhit2}
\< \xi|\e |\xi\> \approx 2 {\left( \frac {m}{2 \pi \hbar^3 T} \right) }^\half 
V_0 \, T' \, h(d,L) \, .
\ee
This is the non-unitary component of the evolution from $t_-$ to $t_+$
when the particle passes through the region $R$ in 1+1 dimensions.  In
3+1 dimensions, this result generalizes to
\be
\< \xi|\e |\xi\> \approx -2 {\left( \frac{m^3}{ \hbar^5 T^3} \right)}^\half 
\frac {V_0 \, {T'}^3 \, h(d,L)}{\pi} \, .
\ee

\end{document}